\documentclass{elsart}

\usepackage{amssymb}
\usepackage{graphicx}

\begin{document}

\begin{frontmatter}



\title{Shell-model Hamiltonian from self-consistent
mean-field model: $N=Z$ nuclei}


\author[a1]{Kazunari~Kaneko,}
\author[a2]{Takahiro~Mizusaki,}
\author[a3,a4]{Yang~Sun,}
\author[a4]{Munetake~Hasegawa}

\address[a1]{Department of Physics, Kyushu Sangyo University,
Fukuoka 813-8503, Japan}
\address[a2]{Institute of Natural Sciences, Senshu University, Tokyo
101-8425, Japan}
\address[a3]{Department of Physics, Shanghai
Jiao Tong University, Shanghai 200240, PR China}
\address[a4]{Institute of Modern Physics, Chinese Academy of
Sciences, Lanzhou 730000, PR China}

\begin{abstract}
We propose a procedure to determine the effective nuclear
shell-model Hamiltonian in a truncated space from a self-consistent
mean-field model, e.g., the Skyrme model. The parameters of pairing
plus quadrupole-quadrupole interaction with monopole force are
obtained so that the potential energy surface of the Skyrme
Hartree-Fock + BCS calculation is reproduced. We test our method for
$N=Z$ nuclei in the $fpg$- and $sd$-shell regions. It is shown that
the calculated energy spectra with these parameters are in a good
agreement with experimental data, in which the importance of the
monopole interaction is discussed. This method may represent a
practical way of defining the Hamiltonian for general shell-model
calculations.
\end{abstract}

\end{frontmatter}


Nuclear structure study is usually carried out with two major groups
of microscopic approaches: the self-consistent mean-field (SCMF)
method \cite{Bender} and the shell model (SM) method \cite{Brown88}.
Both approaches have their advantages and disadvantages. The SCMF
method has a wide applicability across the nuclear chart for global
properties of the ground state, such as the binding energy, nuclear
size, and surface deformation. However, it does not give detailed
spectra of excited states and wave functions. Beyond mean-field
approximations, the angular momentum and particle number projection
method has been applied; but it has been pointed out that there are
some conceptual problems and numerical difficulties
\cite{Dobaczewski07,Bender08}. On the other hand, the SM method has
the advantage that excited energy levels and wave functions are
described properly with many-body correlations included. However, in
the SM approach, the shell model Hamiltonian is required to accord
with each truncated model space, and single-particle energies and
interaction matrix elements must be specific to the mass region. It
is not very clear how to determine these quantities microscopically.
There have been attempts along this line by Brown and Richter
\cite{Brown98} and by Alhassid, Bertsch, and collaborators
\cite{Alhassid1,Alhassid2}. In the former attempt, the SCMF was used
to determine single-particle energies of the SM Hamiltonian, while
in the latter, a procedure for mapping the SCMF onto the SM
Hamiltonian, which includes monopole pairing and
quadrupole-quadrupole ($QQ$) interactions, was proposed. Very
recently, a novel way of determining parameters of the interacting
boson model (IBM) Hamiltonian has been proposed by Nomura {\it et
al.} \cite{Nomura} by using the potential energy surfaces (PES's) of
the SCMF model.

A realistic SM Hamiltonian can in principle be derived from the free
nucleon-nucleon force, and in fact, such microscopic interactions
have been proposed for the $pf$ shell \cite{Kuo,Jensen}. However,
they fail to reproduce excitation spectra, binding energies, and
transitions if many valence nucleons are involved. To overcome this
defect, considerable effort has been put forward on effective
interactions with empirical fit to experimental data
\cite{Brown,Poves,Honma}. On the other hand, realistic effective
interactions in nuclei are expressed in terms of multipole pairing,
multipole particle-hole, and monopole interactions, the dominant
parts of which are the monopole pairing and quadrupole-quadrupole
interactions with monopole terms $(PQQM)$ \cite{Dufour}. This has
actually been confirmed for a wide range of $N \approx Z$ nuclei in
a series of calculations with an extended $PQQM$ interactions
including additional terms (the quadrupole pairing and the
octupole-octupole term) \cite{Hasegawa01,Kaneko02}. This extended
$PQQM$ model has been successfully applied to different nuclei, as
for instance those in the $fp$-shell region \cite{Hasegawa01} and
the $fpg$-shell region \cite{Kaneko02}. The model has only several
parameters, far less than the number of realistic interaction matrix
elements usually contained in shell model calculations. However, its
capability is very much comparable to that of realistic effective
interactions. Thus, the extended $PQQM$ model is not a mere
schematic model, but is a kind of realistic shell model calculation
applicable to a large body of nuclei.

In general, defining an effective SM Hamiltonian, especially for
heavier nuclei where truncation in the shell model space is
necessary, is a very difficult task. It is desired that a SM
Hamiltonian is determined at a more fundamental level, which can not
only locally fit excitation spectra, but also be consistent with a
global description of the ground state properties. It has been
claimed \cite{Dobaczewski} that within the SCMF method, the Skyrme
force contains correct $QQ$ and monopole components, and is able to
describe both low- and high-energy quadrupole excitations. The
Skyrme force including pairing interaction contains $QQ$ and
pairing, as well as monopole components. It is the purpose of the
present Letter that based on the Skyrme SCMF, we propose the
Hamiltonian for the truncated shell model by performing a global PES
mapping. We note that for a shell model using realistic effective
interactions, it may be very difficult to obtain a unique result
when such a global PES mapping is performed because there are too
many interaction matrix elements in the model. However, our $PQQM$
model Hamiltonian has only few parameters, namely, the $g_{0}$,
$\chi$, and monopole strengths (see Eq. (1) below). Therefore, the
$PQQM$ type of interaction is particularly suitable for a global PES
mapping.

\begin{figure}[t]
\includegraphics[totalheight=9.0cm]{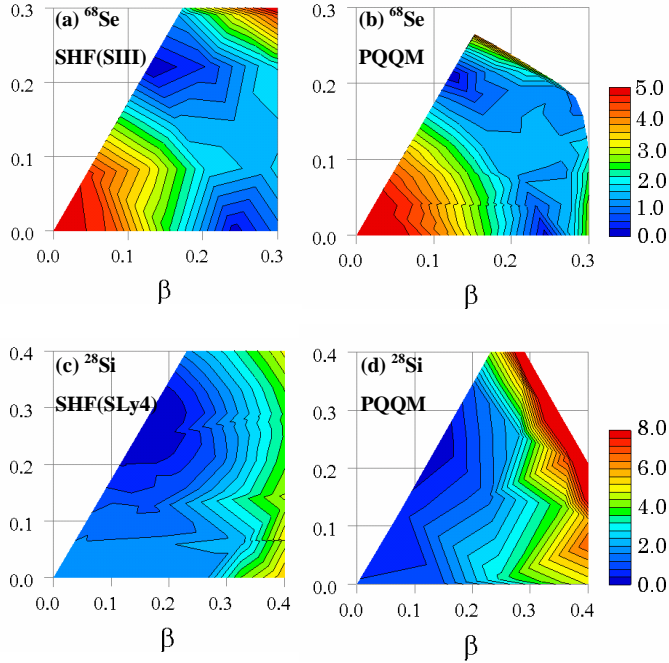}
  \caption{(Color online) PES's for $^{68}$Se and
  $^{28}$Si in the SHF calculation (a and c) and
  the $PQQM$ shell-model calculation (b and d).
  The $PQQM$ parameters are determined so that the $PQQM$ PES
  reproduces approximately that of the SHF.
  Contour spacings are 0.2 MeV and 0.4 MeV for upper and lower
  graphs, respectively.}
  \label{fig1}
\end{figure}

Figure 1a and 1c show PES's on the $\beta$-$\gamma$ plane calculated
by the constrained Skyrme Hartree-Fock + BCS method (hereafter
denoted as SHF), which is imposed by the triaxial degrees of freedom
using the mass quadrupole moments. The plotted energy ranges are up
to 5 MeV for $^{68}$Se and 8 MeV for $^{28}$Si above the respective
energy minimum. For $^{68}$Se, we employ the SIII parameter set
\cite{SIII} of the Skyrme interaction for the mean-field channel,
which has been successful in describing systematically the
ground-state quadrupole deformations in proton- and neutron-rich Kr,
Sr, Zr, and Mo isotopes \cite{Bonche85}. For $^{28}$Si, we use the
SLy4 \cite{SLy4} interaction. We use the ev8 code \cite{ev8} with
pairing interaction of the $\delta$-function type with the strength
$V_{0}$ = 1000 MeV fm$^{3}$. For $^{68}$Se, the long-standing
prediction of a stable oblate deformation was confirmed by the
observation of the oblate ground state band in $^{68}$Se
\cite{Fischer}. Determination of shape was inferred indirectly from
the study of rotational bands, while direct quadrupole measurement
is difficult for these short-lived states. It has been suggested by
various theoretical approaches
\cite{Yamagami,Petrovici,kobayasi,Sun2,Kaneko} that the oblate
configuration coexists with a prolate rotational band, which
constitutes a clear example of oblate-prolate shape coexistence. It
can be seen from Fig. 1 that the PES of the current SCMF calculation
with SIII interaction (Fig. 1a) indeed yields two separate minima at
the oblate and prolate side with deformation $\beta\approx$ 0.24.
For $^{28}$Si, the PES (Fig. 1c) has a minimum at the oblate side
with deformation $\beta\approx$ 0.33, corresponding to the
experimental spectroscopic quadrupole moment $Q_{s}$ = 16
$e$fm$^{2}$.

To connect these SHF results with SM results, we start with the
$PQQM$ model Hamiltonian \cite{Kaneko,Hasegawa}
\begin{eqnarray}
H & = & \sum_{\alpha}\varepsilon_{a}c_{\alpha}^{\dag}c_{\alpha}
-\frac{g_{0}}{2}{P}^{\dag}_{0}\cdot {P}_{0}
-\frac{\chi}{2}{Q}^{\dag}_{2}\cdot {Q}_{2} + V_{\rm m},
\label{eq:1}
\end{eqnarray}
where $\varepsilon_{a}$ is single-particle energy. The second term
in Eq. (\ref{eq:1}) is the monopole pairing interactions with
${P}_{0}$ being the $T=1$, $J=0$ pair operator, and the third term
is the $QQ$ interaction with ${Q}_{2}$ the $T=0$ quadrupole
operator. The last term $V_{\rm m}$ is the monopole force. Due to
isospin-invariance, each of these terms in Eq. (\ref{eq:1}) contains
the {\it p-n} components which play important roles in $N=Z$ nuclei.
The quadrupole-pairing, the octupole-octupole, and the average
monopole terms employed in the previous papers
\cite{Kaneko,Hasegawa} are neglected for simplicity because they do
not affect the current conclusion.

The SM calculation \cite{Kaneko,Hasegawa} is performed by the SM
code \cite{Mizusaki00} for the $fpg$- and $sd$-SM spaces, for which
we assume a closed $^{56}$Ni- and $^{16}$O-core, respectively. Since
the Hamiltonian (\ref{eq:1}) is isospin-invariant, single-particle
energies are taken as the same for protons and neutrons. For the
$fpg$-shell space, the single-particle energies for the $2p_{3/2}$,
$1f_{5/2}$, $2p_{1/2}$, and $1g_{9/2}$ states can be read from the
low-lying states of $^{57}$Ni. We use the experimental values
$\varepsilon_{p3/2}=0.0$, $\varepsilon_{f5/2}=0.77$,
$\varepsilon_{p1/2}=1.11$, and $\varepsilon_{g9/2}=2.50$ (all in
MeV), as in the previous paper \cite{Kaneko}. For the $sd$-shell
space, the single-particle energies for the $1d_{5/2}$, $2s_{1/2}$,
and $1d_{3/2}$ states are employed from USD Hamiltonian
\cite{Brown}. Nuclear shapes including triaxiality are calculated by
the constrained Hartree-Fock (CHF) method \cite{Mizusaki99,Hara99}
and SM PES is defined as the expectation value $\langle H \rangle$
with respect to the CHF state in the $\beta$-$\gamma$ plane.

\begin{figure}[t]
\includegraphics[totalheight=7.0cm]{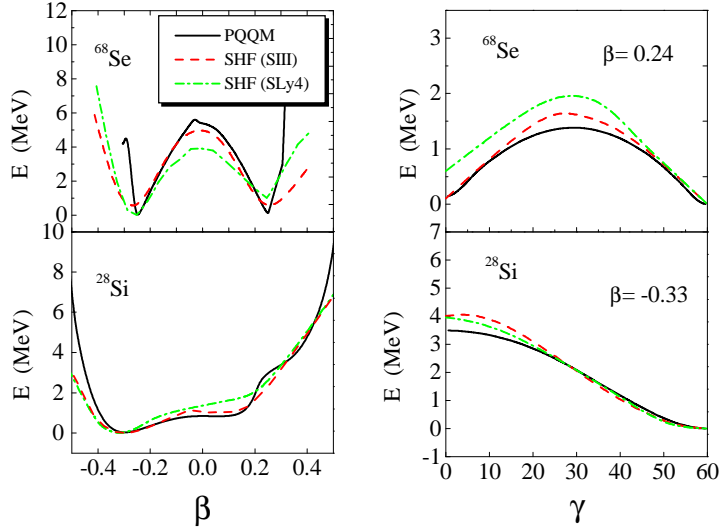}
  \caption{(Color online) PES's for axial and fixed $\beta$
  deformations in $^{68}$Se
  and $^{28}$Si. The left and right panels show PES's along axial
  oblate-to-prolate $\beta$ deformation and variation with triaxiality
  with fixed $\beta$ at the minimum. }
  \label{fig2}
\end{figure}

We now sketch the procedure to determine the pairing, the
quadrupole-quadrupole, and the monopole force strengths by taking
$^{68}$Se and $^{28}$Si as examples. Figure 2 shows the PES's as
functions of axial deformation $\beta$ and of triaxiality $\gamma$
with fixed $\beta$ at the deformation minimum. The PES results in
solid curves are obtained by requiring that the interaction
strengths in the $PQQM$ Hamiltonian are set so as to reproduce the
PES's of the SHF calculation. As one can see, the PES's of the
$PQQM$ calculation reproduce well those of the SHF with SIII for
$^{68}$Se and SLy4 for $^{28}$Si. For large deformations with
$|\beta|>$ 0.24 in $^{68}$Se and $|\beta|>$ 0.4 in $^{28}$Si, the
PES's have the pronounced sharp wall as shown in Fig. 2. This seems
to be a general trend and is probably due to the small truncated
model space. We therefore neglect this sharp wall in the PES
mapping. In this way, the $PQQM$ parameters are uniquely determined.

It is known that the SHF PES pattern depends on the Skyrme
parameterization. To show that the extracted $PQQM$ Hamiltonian has
a general meaning, we present in Fig. 2 also the results for each
nucleus with one more Skyrme interaction, namely the PES's of SHF
with SLy4 for $^{68}$Se and SIII for $^{28}$Si. Comparing the
results, we see that the curves depend only weakly on the choice of
the Skyrme interaction. The essential PES pattern such as shape
coexistence of $^{68}$Se does not alter with a particular
parametrization. This fact has also been realized by the earlier
paper \cite{Nomura}. Here we confirm it for $^{28}$Si and $^{68}$Se
using different Skyrme interactions. For $^{28}$Si, the interaction
strengths in the $PQQM$ Hamiltonian obtained from the Skyrme
interaction SIII are almost the same as those from SLy4. For
$^{68}$Se, the quadrupole interaction has to be modified in order to
fit the SHF-SLy4 PES pattern; however it is only a small reduction
when comparing it with the quadrupole interaction extracted from the
SHF-SIII PES pattern.

The so-obtained $PQQM$ force strengths for the $fpg$-shell space are
$g_{0}=0.270(64/A)$ and $\chi=0.222(64/A)^{5/3}/b^{4}$, with $b$ the
length of harmonic oscillator, and the $T=1$ monopole force strength
is $V_{m}(f_{5/2},p_{1/2};T=1)=-0.25$ MeV. The $PQQM$ Hamiltonian
determined in this way describes quite well the global properties of
these nuclei. In particular, the effect of the monopole shift is
found to be important for producing the oblate minimum. We note that
in the previous paper \cite{Kaneko}, the deformation $\beta$ = 0.20
from the SM calculation with effective charges $e_{\pi}$ = 1.5$e$
and $e_{\nu}$ = 0.5$e$ was smaller than $\beta$ = 0.24 estimated
from the experimental quadrupole moment. Larger effective charges
$e_{\pi}$ = 1.75$e$ and $e_{\nu}$ = 0.75$e$ were therefore needed to
obtain the oblate minimum with $\beta$ = 0.24. Now our new result
for $^{68}$Se with the SM PES calculated from the $PQQM$ Hamiltonian
shows correctly the coexistence of the prolate and oblate minimum at
$|\beta| \approx$ 0.24 (see Fig. 1b).

\begin{figure}[t]
\includegraphics[totalheight=7.0cm]{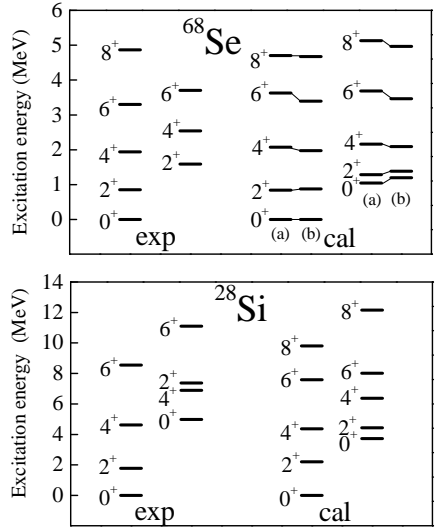}
  \caption{Comparison between experimental and calculated energy levels
  for $^{68}$Se and $^{28}$Si with the $PQQM$ interactions
  determined in the present Letter. In the upper graph for $^{68}$Se,
  the calculated energy levels with the $PQQM$ parameters obtained
  from SIII (marked as (a)) and SLy4 PES's (marked as (b)) are shown
  for the ground-state and side bands.}
  \label{fig3}
\end{figure}

It should be noted that the PES of the standard IBM-2 may not
properly describe triaxial deformation and coexistence of oblate and
prolate shapes because there is no stabilized triaxiality in its
mean field solution, which can be seen from the expectation value of
the IBM Hamiltonian in Eq. (3) of Ref. \cite{Nomura}.

The $PQQM$ PES for the $N=Z$ nucleus $^{28}$Si is shown in Fig. 1d,
which is compared to the SHF results with the SLy4 interaction in
Fig. 1c. The interaction strengths thus-obtained are $g_{0} = 0.50$
and $\chi=4.158A^{-2/3}/b^{4}$, with the monopole interaction
strengths $V_{m}(d_{5/2},d_{3/2};T=1)=-0.20$ and
$V_{m}(s_{1/2},s_{1/2};T=1)=1.0$ MeV. In Fig. 2, the PES's along the
axial deformation and as a function of triaxiality with fixed
minimum $\beta$ are shown. The SM calculation with effective charges
$e_{\pi}$ = 1.5$e$ and $e_{\nu}$ = 0.5$e$ yields a deformation
$\beta = -0.33$ as in the SLy4 PES (see Fig. 2).

\begin{figure}[t]
\includegraphics[totalheight=10cm]{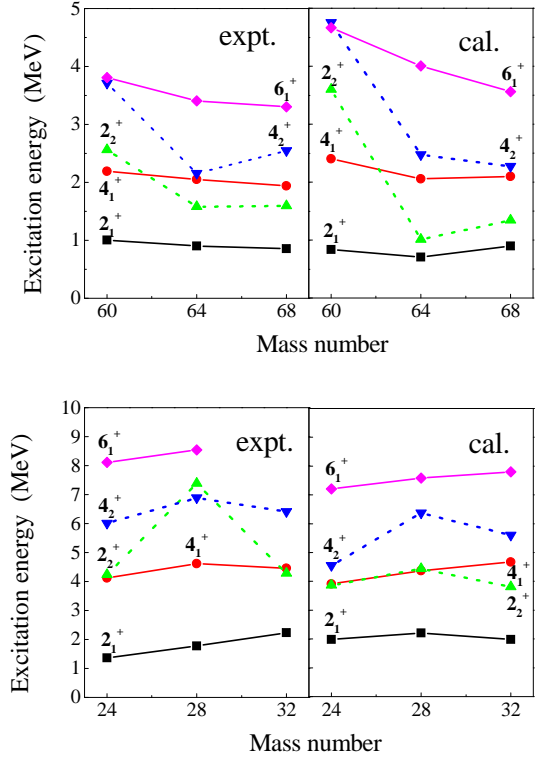}
  \caption{(Color online) Calculated energy levels compared with data for the
  $N=Z$ nuclei in the sd- and fpg-shell regions. The upper and lower panels
  represent results for $^{60}$Zn,
  $^{64}$Ge, $^{68}$Se, and for $^{24}$Mg, $^{28}$Si, $^{32}$S, respectively.}
  \label{fig4}
\end{figure}

In Fig. 3, we compare energy levels between experiment and our SM
calculation for $^{68}$Se and $^{28}$Si, obtained with the $PQQM$
interaction strengths determined from the above procedure. For
$^{68}$Se, we show energy spectra obtained with different PQQM
parameters which are determined from the PES's using the Skyrme
interactions SIII and Sly4. It is seen that the two calculated
energy spectra are resemble each other. Both the experimental ground
and side bands for $^{68}$Se are nicely reproduced. This indicates
that the PQQM Hamiltonian derived from a good PES of the SHF method
works well for producing detailed energy spectra. In Ref.
\cite{Kaneko}, the previous $fpg$-SM calculation for $^{68}$Se using
the phenomenologically-fitted force strengths achieved a reasonable
agreement with data. We note that the $PQQM$ force strengths
proposed in this Letter are close to those fitted ones in
\cite{Kaneko}. The calculation for $^{68}$Se predicts the first
excited $0_{2}^{+}$ state. Our analysis for quadrupole moments
indicates that the ground-state has an oblate deformation and the
side band has a prolate shape. For $^{28}$Si, the calculated ground
band reproduces the data well. The calculation indicates the side
band built on the first excited $0_{2}^{+}$ state; however it does
not exhibit the inversion of the second $2_{2}^{+}$ and $4_{2}^{+}$
states as suggested by the current data.

Next we test this procedure with the neighboring $N=Z$ nuclei of
$^{68}$Se and $^{28}$Si. Figure 4 shows a comparison of our
calculated energy levels with data for $^{60}$Zn, $^{64}$Ge, and
$^{68}$Se, and for $^{24}$Mg, $^{28}$Si, and $^{32}$S. The
calculation correctly reproduces the trend of level variation as
mass number changes, with only one exception in the second excited
$2_{2}^{+}$ state of $^{28}$Si, as mentioned before.

$E2$ transition probabilities for the positive-parity yrast and
excited states in $^{28}$Si and $^{68}$Se are shown in Table I. For
$^{28}$Si, our calculated $B(E2)$ values are in good agreement with
the experimental data. For $^{68}$Se, the quadrupole deformation
obtained from our calculated $B(E2;2_{1}^{+}\rightarrow 0_{1}^{+})$
is $\beta\sim$ 0.26, which is consistent with the experimental
estimation $\beta\sim$ 0.27 by Fischer {\it et al.} \cite{Fischer}
and 0.30 by Jenkins {\it et al.} \cite{Jenkins}. In Table I, we list
also the theoretical $B(E2)$ values by Petrovici {\it et al.}
\cite{Petrovici} with the Excited Vampir calculation. Their
estimated deformation is $\beta\sim$ 0.37, which is much larger than
ours, and inconsistent with the experimental estimation.

\begin{figure}[t]
\includegraphics[totalheight=9cm]{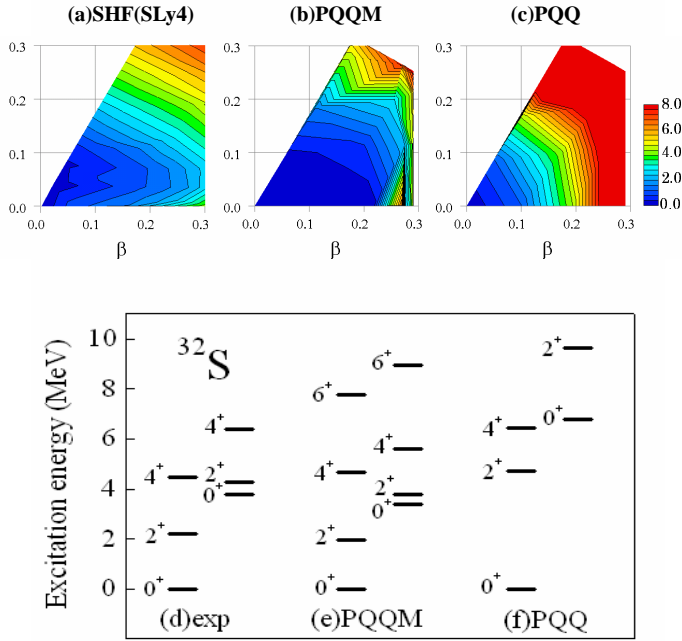}
  \caption{(Color online) PES's and energy levels for $^{32}$S.
  (a) PES of the SHF(SLy4), (b) PES of the $PQQM$ model,
  (c) PES of the $PQQ$ model (without monopole interactions). In the
  lower plot, the SM energy levels of $PQQM$ in (e) and $PQQ$ in (f)
  are compared with experimental data in (d).
  Contour spacings in a, b and c are 0.4 MeV.}
  \label{fig5}
\end{figure}

\begin{table}[t]
\caption{Calculated $B(E2)$ values for positive-parity yrast and
excited states in $^{28}$Si and $^{68}$Se, which are compared with
the known experimental ones for $^{28}$Si and the theoretical values
of Petrovici {\it et al.} \cite{Petrovici} for $^{68}$Se,
respectively.}
\begin{tabular*}{85mm}{@{\extracolsep{\fill}}ccccc} \hline\hline
        & \multicolumn{2}{c}{$^{28}$Si  [$e^2$fm$^{4}$]}
        & \multicolumn{2}{c}{$^{68}$Se  [$e^2$fm$^{4}$]}    \\ \hline
$I_i^\pi \rightarrow I_f^\pi$ & Expt. & Calc. & Petrovici {\it et al.}  & Calc. \\ \hline
$2_1^+ \rightarrow 0_1^+$       &  66.7  &  55.7  &   966   &  503.3   \\
$4_1^+ \rightarrow 2_1^+$       &  69.7  &  51.2  &  1381   &  609.6   \\
$6_1^+ \rightarrow 4_1^+$       &  50.0  &  54.5  &  1402   &  594.6   \\
$8_1^+ \rightarrow 6_1^+$       &        &  34.9  &  1710   &          \\
$0_2^+ \rightarrow 2_1^+$       &  43.4  &  99.7  &         &          \\
$2_2^+ \rightarrow 0_2^+$       &        &        &         &  511.7   \\
$4_2^+ \rightarrow 2_2^+$       &        &        &         &  553.8   \\
$6_2^+ \rightarrow 4_2^+$       &        &        &         &  495.4   \\
$8_2^+ \rightarrow 6_2^+$       &        &        &         &   44.1   \\ \hline\hline
\end{tabular*}
\label{table1}
\end{table}

We take $^{32}$S as an example to discuss the monopole
effects on PES. The monopole interaction
$V_{m}(d_{5/2},d_{3/2};T=1)$ between the spin-orbit partners
$d_{5/2}$ and $d_{3/2}$ is known to be very important for the
$sd$-shell spectra. As the Fermi energy approaches the $d_{3/2}$
orbit, the monopole interactions $V_{m}(d_{5/2},d_{3/2};T=1)$ and
$V_{m}(s_{1/2},s_{1/2};T=1)$ act on the relevant orbits and affect
both PES and energy levels. In Fig. 5, the PES's and energy levels
in the SM calculation with and without the monopole force are
respectively compared with the SHF PES's and with experimental
energy levels. The energy range is up to 8 MeV above the energy
minimum. Figure 5b exhibits PES's of the $PQQM$ model calculated
with the determined parameters by comparison with the PES's of the
SLy4 interaction in Fig. 5a. The calculated energy levels are shown
in Fig. 5e, which are compared with data in Fig. 5d. As can be seen,
the $PQQM$ calculation with the present interaction strengths
reproduces data well. To see the monopole effects on PES and energy
levels, we switch off all the monopole interactions and show the
results in Fig. 5c and 5f. One sees that the PES in Fig. 5c does not
reproduce that of Fig. 5a, and the calculated energy levels in Fig.
5f lie too high when compared with data. We thus conclude that the
monopole force $V_{m}$ is important for a correct reproduction of
both the SHF PES and experimental energy levels in $^{32}$S.

\begin{figure}[t]
\includegraphics[totalheight=6cm]{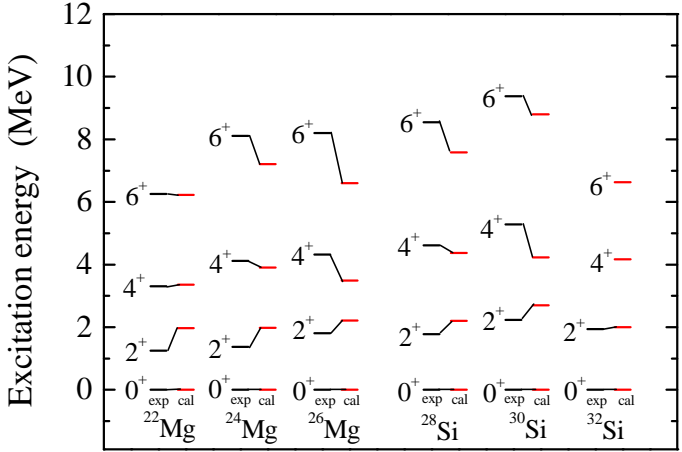}
  \caption{(Color online) Calculated energy levels compared with experimental data
   for the Mg and Si isotopic chains.}
  \label{fig6}
\end{figure}

Finally, we show in Fig. 6 a systematical comparison between theory
and experiment for the energy levels along the Mg and Si isotopic
chains. We can see that the calculated energy levels for the
low-lying $2^{+}$ and $4^{+}$ states reproduce fairly well those of
the experimental data, while the $6^{+}$ states lie a little higher
than experiment.

To summarize, for a correct SM description of nuclear spectra
phenomenologically-adjusted effective interactions are usually
introduced. In the present Letter, we have presented a procedure to
define the SM Hamiltonian for a truncated space at a more
fundamental level, by performing a global PES mapping with the SCMF
results of the Skyrme interaction. The parameters of the $PQQM$
model have been determined so as to reproduce the overall pattern of
the PES of the SHF calculation. The $PQQM$ SM calculations with the
determined forces have reproduced well the experimental energy
levels for the $N=Z$ nuclei in the $fpg$- and $sd$-shell regions.
Effects brought by the monopole interactions have been discussed.
This work may represent a practical method of defining SM
Hamiltonian from microscopic mean-field theories, and therefore may
have general applications in other shell models. For example, the
Projected Shell Model \cite{PSM} that employs the separable forces
can adopt this method.

In the present work, single-particle energies in the SM calculation
are taken from experiment as usual. For consistency, we should have
used single-particle energies of the SCMF. However, it is well known
that the SHF single-particle energies cannot be directly compared
with experimental data. A recent study \cite{Zalewski} suggests that
fitting the spin-orbit and tensor parts of the SCMF to the
spin-orbit splittings improves considerably the single-particle
properties of the SCMF. Therefore, there are two possibilities for
our choice of single-particle states. One is to use the experimental
single-particle energies as in the present work, and the other is to
use the improved SCMF single-particle energies that include the
tensor interaction. The latter deserves more investigation, and will
be our future goal of study. Application of the present method to
neutron-rich nuclei including the tensor interaction in the SHF is
also in progress.

YS was supported by the National Natural Science Foundation of China
under contract 10875077 and by the Chinese Major State Basic
Research Development Program through grant 2007CB815005.



\end{document}